\documentclass[preprint,authoryear,12pt]{elsarticle}

\usepackage{amssymb}
\usepackage{graphicx}
\usepackage{amsmath}

\journal{Computer Speech and Language}

\begin{document}

\begin{frontmatter}



\title{A Comparative Study of Glottal Source Estimation Techniques}


\author[label1]{Thomas Drugman}
\author[label2]{Baris Bozkurt}
\author[label1]{Thierry Dutoit}
\address[label1]{TCTS Lab, University of Mons, 31 Boulevard Dolez, 7000 Mons, Belgium}
\address[label2]{Department of Electrical \& Electronics Engineering, Izmir Institute of Technology, Gulbahce Koyu 35430, Urla, Izmir, Turkey}

\begin{abstract}
Source-tract decomposition (or glottal flow estimation) is one of the basic problems of speech processing. For this, several techniques have been proposed in the literature. However studies comparing different approaches are almost nonexistent. Besides, experiments have been systematically performed either on synthetic speech or on sustained vowels. In this study we compare three of the main representative state-of-the-art methods of glottal flow estimation: closed-phase inverse filtering, iterative and adaptive inverse filtering, and mixed-phase decomposition. These techniques are first submitted to an objective assessment test on synthetic speech signals. Their sensitivity to various factors affecting the estimation quality, as well as their robustness to noise are studied. In a second experiment, their ability to label voice quality (tensed, modal, soft) is studied on a large corpus of real connected speech. It is shown that changes of voice quality are reflected by significant modifications in glottal feature distributions. Techniques based on the mixed-phase decomposition and on a closed-phase inverse filtering process turn out to give the best results on both clean synthetic and real speech signals. On the other hand, iterative and adaptive inverse filtering is recommended in noisy environments for its high robustness.
\end{abstract}


\end{frontmatter}



\section{Introduction}\label{sec:Intro}
%
%
%
%
Speech results from filtering the glottal flow by the vocal tract cavities, and converting the resulting velocity flow into pressure at the lips \citep{Quatieri}. In many speech processing applications, it is important to separate the contributions from the glottis and the vocal tract. Achieving such a \emph{source-filter deconvolution} could lead to a distinct characterization and modeling of these two components, as well as to a better understanding of the human phonation process. Such a decomposition is thus a preliminary condition for the study of glottal-based vocal effects, which can be segmental (as for vocal fry), or be controlled by speakers on a separate, supra-segmental layer (as it is the case for the voice quality modifications mentioned in Section \ref{sec:Real}). Their dynamics is very different from that of the vocal tract contribution, and requires further investigation. Glottal source estimation is then a fundamental problem in speech processing, finding applications in speech synthesis \citep{Cabral}, voice pathology detection \citep{Drugman-Patho}, speaker recognition \citep{Plumpe}, emotion analysis/synthesis \citep{Alku-Emotion}, etc.

In this paper, we limit our scope to the methods which perform an estimation of the glottal source contribution directly from the speech waveform. Although some devices such as electroglottographs or laryngographs, which measure the impedance between the vocal folds (but not the glottal flow itself), are informative about the glottal behaviour \citep{EGG}, in most cases the use of such aparatus is inconvenient and only the speech signal is available for analysis. This problem is then a typical case of blind separation, since neither the vocal tract nor the glottal contribution are observable. This also implies that no quantitative assessment of the performance of glottal source estimation techniques is possible on natural speech, as no target reference signal is available.

As one of the basic problems and challenges of speech processing research, glottal flow estimation has been studied by many researchers and various techniques are available in the literature \citep{Walker}. However the diversity of algorithms and the fact that the reference for the actual glottal flow is not available often leads to the questionability about relative effectiveness of the methods in real life applications. In most of studies, tests are performed either on synthetic speech or on a few recorded sustained vowels. In addition, very few comparative studies exist (such as \citep{Sturmel}). In this paper, we compare three of the main representative state-of-the-art methods: closed-phase inverse filtering, iterative and adaptive inverse filtering, and mixed-phase decomposition. For testing, we first follow the common approach of using a large set of synthetic speech signals (by varying synthesis parameters independently), and then we examine how these techniques perform on a large real speech corpus. In the synthetic speech tests, the original glottal flow is available, so that objective measures of decomposition quality can be computed. In real speech tests the ability of the methods to discriminate different voice qualities (tensed, modal and soft) is studied on a large database (without limiting data to sustained vowels).

The paper is structured as follows. In Section \ref{sec:Estimation} the main state-of-the-art methods for glottal source estimation are reviewed, and the three techniques compared in this study are detailed. Section \ref{sec:Parametrization} discusses how the resulting glottal signal can be parametrized both in time and frequency domains. The three methods are evaluted in Section \ref{sec:Synthetic} through a wide systematic study on synthetic signals. Their robustness to noise, as well as the impact of the various factors that may affect source-tract separation, are investigated. Section \ref{sec:Real} presents decomposition results on a real speech database containing various voice qualities, and shows that the glottal source estimated by the techniques considered in this work conveys relevant information about the phonation type. Finally Section \ref{sec:conclu} draws the conclusions of this study.

\section{Glottal Source Estimation}\label{sec:Estimation}

\begin{figure}[!ht]
  \centering
  \includegraphics[width=0.45\textwidth]{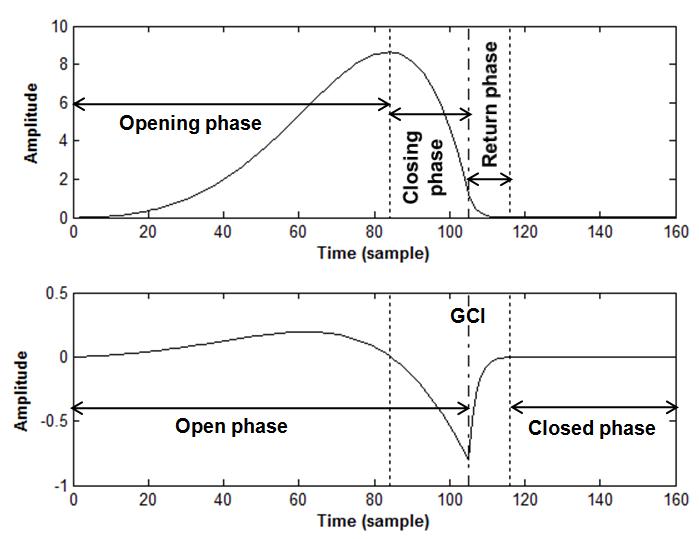}
  \caption{Typical waveforms, according to the Liljencrants-Fant (LF) model, of one cycle of: (top) the glottal flow, (bottom) the glottal flow derivative. The various phases of the glottal cycle, as well as the Glottal Closure Instant (GCI) are also indicated.}
  \label{fig:LFmodel}
\end{figure}

Glottal flow estimation mainly refers to the estimation of the voiced excitation of the vocal tract. During the production of voiced sounds, the airflow arising from the trachea causes a quasi-periodic vibration of the vocal folds \citep{Quatieri}, organized into so-called opening/closure cycles. During the \emph{open phase}, vocal folds are progressively displaced from their initial state due to the increasing subglottal pressure. When the elastic displacement limit is reached, they suddenly return to this position during the so-called \emph{return phase}. Figure \ref{fig:LFmodel} displays the typical shape of one cycle of the glottal flow (Fig.\ref{fig:LFmodel}(a)) and its time derivative (Fig.\ref{fig:LFmodel}(b)) according to the Liljencrants-Fant (LF) model \citep{LF}. It is often prefered to gather the lip radiation effect (whose action is close to a differentiation operator) with the glottal component, and work in this way with the glottal flow derivative on the one hand, and with the vocal tract contribution on the other hand. It is seen in Figure \ref{fig:LFmodel} (bottom plot) that the boundary between open and return phases corresponds to a particular event called the Glottal Closure Instant (GCI). GCIs refer to the instances of significant excitation of the vocal tract \citep{Drugman-GCI}. Being able to determine their location is of particular importance in so-called pitch-synchronous speech processing techniques, and in particular for a more accurate separation between vocal tract and glottal contributions.

The main techniques for estimating the glottal source directly from the speech waveform are now reviewed. Relying on the speech signal alone, as it is generally the case in real applications, allows to avoid the use of intrusive (e.g video camera at the vocal folds) or inconvenient (e.g. laryngograph) device.

Such techniques can be separated into two classes, according to the way they perform the source-filter separation. The first category (Section \ref{ssec:InvFilt}) is based on inverse filtering, while the second one (Section \ref{ssec:MixedPhase}) relies on the mixed-phase properties of speech.

\subsection{Methods based on Inverse Filtering}\label{ssec:InvFilt}

Most glottal source estimation techniques are based on an inverse filtering process. These methods first estimate a parametric model of the vocal tract, and then obtain the glottal flow by removing the vocal tract contribution via inverse filtering. The methods in this category differ by the way the vocal tract is estimated. In Section \ref{sssec:CPIF} this estimation is computed during the glottal closed phase, while in Section \ref{sssec:IAIF} an iterative and/or adaptive procedure is used. A more extended review of the inverse filtering-based process for glottal waveform analysis can be found in \citep{Walker}.

\subsubsection{Closed Phase Inverse Filtering}\label{sssec:CPIF}

Closed phase refers to the timespan during which the glottis is closed (see Figure \ref{fig:LFmodel}). During this period, the effects of the subglottal cavities are minimized, providing a better way for estimating the vocal tract transfer function. Therefore, methods based on a Closed Phase Inverse Filtering (CPIF) estimate a parametric model of the spectral envelope, computed during the estimated closed phase duration \citep{Wong}. The main drawback of these techniques lies in the difficulty in obtaining an accurate determination of the closed phase. Several approaches have been proposed in the literature to solve this problem. In \citep{Veeneman}, authors use information from the electroglottographic signal (which is avoided in this study) to identify the period during which the glottis is closed. In \citep{Plumpe}, it was proposed to determine the closed phase by analyzing the formant frequency modulation between open and closed phases. In \citep{Alku-CP}, the robustness of CPIF to the frame position was improved by imposing some dc gain constraints. Besides this problem of accurate determination of the closed phase, it may happen that this period is so short (for high-pitched voices) that not enough samples are available for a reliable filter estimation. It was therefore proposed in \citep{Brookes} a technique of multicycle closed-phase LPC, where a small number of neighbouring glottal cycles are considered in order to have enough data for an accurate vocal tract estimation. Finally note that an approach allowing non-zero glottal wave to exist over closed glottal phases was proposed in \citep{Deng}.

In this study, the CPIF-based technique that is used is based on a Discrete All Pole (DAP, \citep{ElJaroudi1991}) inverse filtering process estimated during the closed phase. In order to provide a better fitting of spectral envelopes from discrete spectra \citep{ElJaroudi1991}, the DAP technique aims at computing the parameters of an autoregressive model by minimizing a discrete version of the Itakura-Saito distance \citep{Itakura}, instead of the time squared error used by the traditional LPC. The use of the Itakura-Saito distance is justified as it is a spectral distortion measure arising from the human hearing perception. The closed phase period is determined using the Glottal Opening and Closure Instants (GCIs and GOIs) located by the algorithm detailed in \citep{Drugman-GCI}. This algorithm has been shown to be effective for reliably and accurately determining the position of these events on a large corpus containing several speakers. For tests with synthetic speech, the exact closed phase period is known and is used for CPIF. Note that for high-pitched voices, two analysis windows were used as suggested in \citep{Brookes}, \citep{Yegna1} and \citep{Plumpe}. In the rest of the paper, speech signals sampled at 16 kHz are considered, and the order for DAP analysis is fixed to 18 (=$F_s/1000+2$, as commonly used in the literature). Through our experiments, we reported that the choice of the DAP order is not critical in the usual range, and that working with an order comprised between 12 and 18 leads to sensibly similar results.

\subsubsection{Iterative and/or Adaptive Inverse Filtering}\label{sssec:IAIF}

Some methods are based on iterative and/or adaptive procedures in order to improve the quality of the glottal flow estimation. In \citep{ARXLF}, Fu and Murphy proposed to integrate, within the AutoRegressive eXogenous (ARX) model of speech production, the LF model of the glottal source. The resulting  ARXLF model is estimated via an adaptive and iterative optimization \citep{Vincent}. Both source and filter parameters are consequently jointly estimated. The method proposed by Moore in \citep{Moore} iteratively finds the best candidate for a glottal waveform estimate within a speech frame, without requiring a precise location of the GCIs. Finally a popular approach was proposed by Alku in \citep{IAIF} and called Iterative Adaptive Inverse Filtering (IAIF). This method is based on an iterative refinement of both the vocal tract and the glottal components. In \citep{DAP}, the same authors proposed an improvement, in which the LPC analysis is replaced by the Discrete All Pole (DAP) modeling technique \citep{ElJaroudi1991}, shown to be more accurate for high-pitched voices.

As a representative technique of this category, the IAIF method proposed by Alku in \citep{IAIF} is considered in the rest of this paper. More precisely, we used the implementation of the IAIF method \citep{Airas2008} from the toolbox available on the TKK Aparat website \citep{Aparat}, with its default options.



\subsection{Mixed-Phase Decomposition}\label{ssec:MixedPhase}

The methods presented in this Section rely on the mixed-phase model of speech \citep{MixedPhase}. According to this model, speech is composed of both minimum-phase (i.e causal) and maximum-phase (i.e anticausal) components. While the vocal tract impulse response and the glottal \emph{return phase} of the glottal component can be considered as minimum-phase signals, it has been shown in \citep{Doval-CALM} that the glottal \emph{open phase} of the glottal flow is a maximum-phase signal. Besides it has been shown in \citep{Gardner} that mixed-phase models are appropriate for modeling voiced speech due to the maximum-phase nature of the glottal excitation. They showed that the use of an anticausal all-pole filter for the glottal pulse is necessary to resolve magnitude and phase information correctly. The key idea of mixed-phase decomposition methods is then to separate minimum from maximum-phase components of speech, where the latter is only due to the glottal contribution.

A crucial issue in mixed-phase separation is the weighting window that is applied to the speech signal for short-term analysis. Indeed, since the decomposition is based on phase properties, windowing may have a dramatic influence. It has been shown that GCI-synchronization, as well as the respect of some constraints on the window length and function, are essential for guaranteeing a correct decomposition \citep{CCD}, \citep{ZZT}. Throughout the rest of this study, we use an appropriate GCI-centered two pitch period-long Blackman window satisfying these conditions.

In previous works, we proposed two approaches achieving such a decomposition: a technique based on the Zeros of the Z-Transform (ZZT, \citep{ZZT}), and one based on the Complex Cepstrum Decomposition (CCD, \citep{CCD}, \citep{Quatieri}). Both techniques are briefly presented in Sections \ref{sssec:ZZT} and \ref{sssec:CCD} and depicted in Figure \ref{fig:MixedPhase}. Finally, the methods are shown to be functionnaly equivalent in Section \ref{sssec:Equivalence}.

\subsubsection{Zeros of the Z-Transform (ZZT)}\label{sssec:ZZT}

For a series of $N$ samples $(x(0),x(1),...,x(N-1))$ taken from a
discrete signal $x(n)$, the $ZZT$ representation is defined as
the set of roots (zeros) $(Z_1,Z_2,...Z_{N-1})$ of the corresponding
z-transform $X(z)$:

\begin{align}
X(z)&=\sum_{n=0}^{N-1} x(n)z^{-n}\label{eq:ZZT1}\\
&=x(0)z^{-N+1}\prod_{m=1}^{N-1} (z-Z_m)\label{eq:ZZT2}\\
&=x(0)z^{-N+1}\prod_{k=1}^{M_o} (z-Z_{max,k}) \prod_{k=1}^{M_i} (z-Z_{min,k})\label{eq:ZZT3}
\end{align}

\begin{figure*}[!ht]
  \centering
  \includegraphics[width=0.95\textwidth]{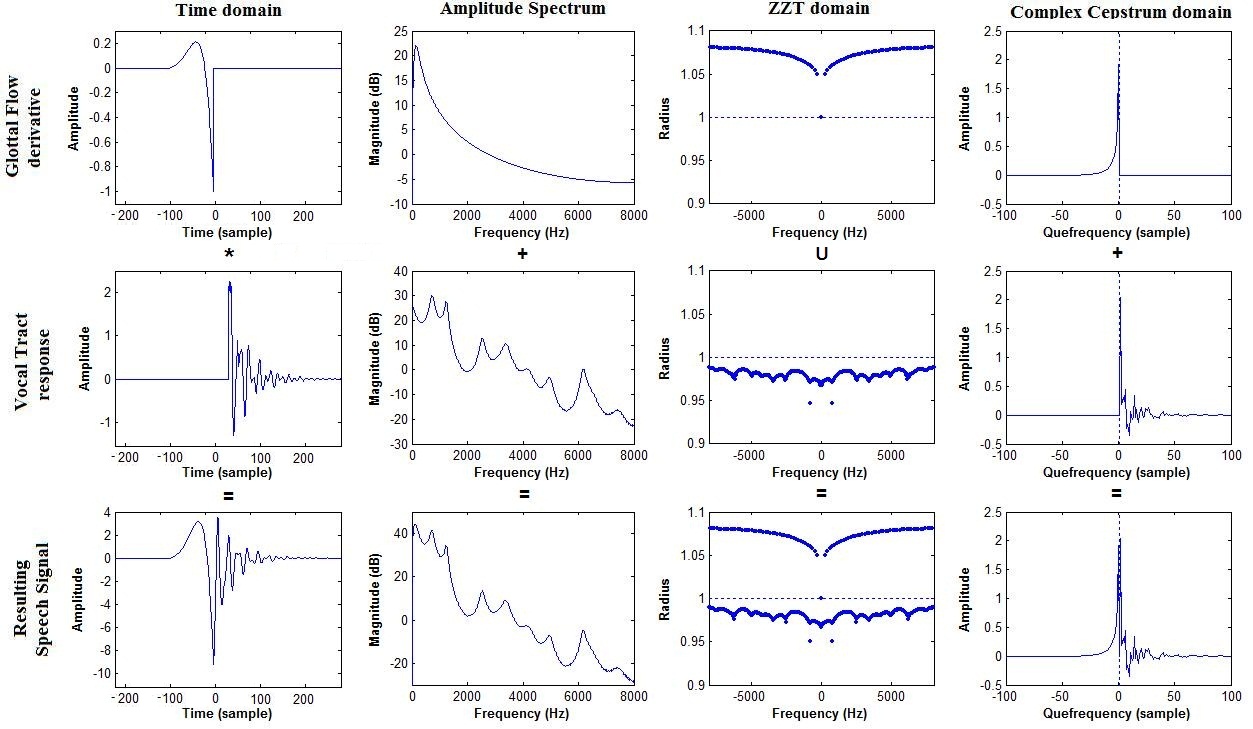}
  \caption{Illustration of mixed-phase decomposition. Rows respectively exhibit the following signals: the glottal flow derivative (top), the vocal tract response (middle), and the resulting speech signal (bottom). Each column corresponds to a domain of representation of these signals: time domain (first column), amplitude spectrum (second column), ZZT representation in polar coordinates (third column), and complex cepstrum domain (fourth column). Interestingly, convolution in the time domain corresponds to the union operator in the ZZT domain and to the addition operator in the complex cepstrum domain. The ZZT and CC domains are suited for achieving mixed-phase decomposition since minimum and maximum-phase components become linearly separable. In the ZZT domain, the unit circle is used as a discriminant boundary, while the quefrency origin is used as a boundary in the complex cepstrum domain.}
  \label{fig:MixedPhase}
\end{figure*}

To achieve the ZZT-based decomposition of speech, speech frames are first weighted by a specific window (see above). When computing the ZZT of this signal as in Equation \ref{eq:ZZT3}, some roots $Z_{max,k}$ fall outside the unit circle. These are due to the maximum-phase (i.e anticausal) component of speech, and are consequently only related to the glottal open phase. On the opposite, roots located inside the unit circle $Z_{min,k}$ are due to the minimum-phase component of speech, i.e mainly to the vocal tract impulse response. Mixed-phase decomposition can then be easily achieved in the ZZT domain, using the unit circle as a discriminant boundary (see Figure \ref{fig:MixedPhase}, third column).

\subsubsection{Complex Cepstrum Decomposition (CCD)}\label{sssec:CCD}

The Complex Cepstrum (CC) $\hat{x}(n)$ of a discrete signal $x(n)$ is defined by the following equations \citep{Oppenheim}:

\begin{equation}\label{eq:DFT}
X(\omega)=\sum_{n=-\infty}^{\infty} x(n)e^{-j\omega n}
\end{equation}

\begin{equation}\label{eq:ComplexLog}
\log[X(\omega)]=\log(|X(\omega)|)+j\angle{X(\omega)}
\end{equation}

\begin{equation}\label{eq:ComplexCepstrum}
\hat{x}(n)=\frac{1}{2\pi}\int_{-\pi}^{\pi}{\log[X(\omega)]e^{j\omega n}}\emph{d}\omega
\end{equation}

where Equations \ref{eq:DFT}, \ref{eq:ComplexLog}, \ref{eq:ComplexCepstrum} respectively correspond to a Discrete-Time Fourier Transform (DTFT), a complex logarithm and an inverse DTFT (IDTFT). Decomposition in the CC domain arises from the fact that the complex cepstrum $\hat{x}(n)$ of an anticausal (causal) signal is zero for all $n$ positive (negative). Retaining only the negative indexes of the CC makes then it possible to estimate the glottal contribution. The separation in the complex cepstrum domain using the quefrency origin as a discriminant boundary is clearly seen in Figure \ref{fig:MixedPhase}, fourth column.

%

\subsubsection{Equivalence between ZZT and CCD}\label{sssec:Equivalence}

If $X(z)$ is written as in Equation \ref{eq:ZZT3}, it can be easily shown that the corresponding complex cepstrum can be expressed as \citep{Oppenheim}:

\begin{equation}\label{eq:DevComplCepstrum}
\hat{x}(n)= \left\{
\begin{array}{ll}
|x(0)| & \emph{for }n=0\\
\sum_{k=1}^{M_o}{\frac{{Z_{max,k}}^n}{n}} & \emph{for } n < 0\\
\sum_{k=1}^{M_i}{\frac{{Z_{min,k}}^n}{n}} & \emph{for } n > 0
\end{array} \right.
\end{equation}

This equation shows the narrow link between the ZZT and the CCD techniques. These two methods can then be seen as two different ways of performing the same operation: separate the minimum and maximum-phase components from a given z-transform $X(z)$. Nevertheless, although functionnaly equivalent, it has been shown \citep{Pedersen}, \citep{CCD} that CCD performs much faster (speed is increased between 10 and 100 times for a sampling rate of 16 kHz, depending on the pitch period). This may be explained by the fact that it only relies on FFT and IFFT operations while ZZT requires the factoring of high-order polynomials.

As a method achieving mixed-phase separation, the CCD is considered in the rest of this paper for its higher computational speed. To guarantee good mixed-phase properties \citep{CCD}, GCI-centered two pitch period-long Blackman windows are used. For this, GCIs were located on real speech using the technique we proposed in \citep{Drugman-GCI}. CC is calculated as explained in Section \ref{sssec:CCD} and FFT is computed on a sufficiently large number of points (typically 4096), which facilitates phase unwrapping.

\section{Glottal Source Parametrization}\label{sec:Parametrization}

Once the glottal signal has been estimated by any of the aforementioned algorithms, it is interesting to derive a parametric representation of it, using a small number of parameters. Various approaches, both in the time and frequency domains, have been proposed to characterize the human voice source. This Section gives a brief overview of the most commonly used parameters in the literature, since some of them are used in Sections \ref{sec:Synthetic} and \ref{sec:Real}.

\subsection{Time-domain features}\label{ssec:TimeParam}
Several time-domain features can be expressed as a function of time intervals derived from the glottal waveform \citep{Alku-Time}. These are used to characterize the shape of the waveform, by capturing for example the location of the primary or secondary opening instant \citep{QOQ}, of the glottal flow maximum, etc. The formulation of the source signal in the commonly used LF model \citep{LF} is based on time-domain parameters, such as the Open Quotient $O_q$, the Asymmetry coefficient $\alpha_m$, or the Voice Speed Quotient $S_q$ \citep{Doval-Spectrum}. However in most cases these instants are difficult to locate with precision from the glottal flow estimation. Avoiding this problem and prefered to the traditional Open Quotient, the Quasi-Open Quotient (QOQ) was proposed as a parameter describing the relative open time of the glottis \citep{Hacki}. It is defined as the ratio between the quasi-open time and the quasi-closed time of the glottis, and corresponds to the timespan (normalized to the pitch period) during which the glottal flow is above $50\%$ of the difference between the maximum and minimum flow. Note that QOQ was used in \citep{QOQ} for studying the physical variations of the glottal source related to the vocal expression of stress and emotion. In \citep{Airas} various variants of $Oq$ have been tested in terms of the degree by which they reflect phonation changes. QOQ was found to be the best for this task.

Another set of parameters is extracted from the amplitude of peaks in the glottal pulse or its derivative \citep{Gobl}. The Normalized Amplitude Quotient (NAQ) proposed by Alku in \citep{Alku-NAQ} turns out to be an essential glottal feature. NAQ is a parameter characterizing the glottal closing phase \citep{Alku-NAQ}. It is defined as the ratio between the maximum of the glottal flow and the minimum of its derivative, normalized with respect to the fundamental period. Its robustness and efficiency to separate different types of phonation was shown in \citep{Alku-NAQ}, \citep{Airas}. Note that a quasi-similar feature, called \emph{basic shape parameter}, was proposed by Fant in \citep{Fant-Revisited}, where it was qualified as \emph{"most effective single measure for describing voice qualities"}.

In \citep{Plumpe}, authors propose to use 7 LF parameters and 5 energy coefficients (defined in 5 subsegments of the glottal cycle) respectively for characterizing the coarse and fine structures of the glottal flow estimation. Finally some approaches aim at fitting a model on the glottal flow estimate by computing a distance in the time domain \citep{Plumpe}, \citep{Drugman-Eusipco08}.

\subsection{Frequency-domain features}\label{ssec:FreqParam}
In the frequency domain, the LF model presents a low-frequency resonance called the \emph{glottal formant} \citep{Doval-Spectrum} (see the amplitude spectrum of the glottal flow derivative in Figure \ref{fig:MixedPhase}, row 1, column 2). Some approaches characterize the glottal formant both in terms of frequency and bandwidth \citep{CCD}. By defining a spectral error measure, other studies try to match a model to the glottal flow estimation \citep{SpecFit}, \citep{Fant-Revisited}, \citep{Drugman-Eusipco08}. This is also the case for the Parabolic Spectrum Parameter (PSP) proposed in \citep{PSP}.

An extensively used measure is the $H1-H2$ parameter \citep{Fant-Revisited}. This parameter is defined as the ratio between the amplitudes of the magnitude spectrum of the glottal source at the fundamental frequency and at the second harmonic \citep{Klatt}, \citep{Titze}. It has been widely used as a measure characterizing voice quality \citep{Hansson}, \citep{Fant-Revisited}, \citep{Alku-CP}.

For quantifying the amount of harmonics in the glottal source, the Harmonic to Noise Ratio (HNR) and the Harmonic Richness Factor (HRF) have been proposed in \citep{HNR} and \citep{HRF}. More precisely, HRF quantifies the amount of harmonics in the magnitude spectrum of the glottal source. It is defined as the ratio between the sum of the amplitudes of harmonics, and the amplitude at the fundamental frequency \citep{Childers}. It was shown to be informative about the phonation type in \citep{HRF} and \citep{Alku-CP}.

\section{Experiments on Synthetic Speech}\label{sec:Synthetic}

The first experimental protocol we opted for is close to the one presented in \citep{Sturmel}. Decomposition is achieved on synthetic speech signals (sampled at 16 kHz) for various test conditions. The idea is to cover the diversity of configurations one can find in continuous speech by varying all parameters over their whole range. Synthetic speech is produced according to the source-filter model by passing a known sequence of Liljencrants-Fant (LF) glottal waveforms \citep{LF} through an auto-regressive filter extracted by LPC analysis (with an order of 18) from real sustained vowels uttered by a female speaker. As the mean pitch during these utterances was about 180 Hz, it can be considered that fundamental frequency should not exceed 100 and 240 Hz in continuous speech. For the LF parameters, the Open Quotient $Oq$ and Asymmetry coefficient $\alpha_m$ are varied through their common range (see Table \ref{tab:Range}). For the filter, 14 types of typical vowels are considered. Noisy conditions are modeled by adding a white Gaussian noise to the speech signal, from almost clean conditions ($SNR = 80dB$) to strongly adverse environments ($SNR = 10dB$). Table \ref{tab:Range} summarizes all test conditions, which makes a total of slightly more than 250,000 experiments. It is worth mentioning that the synthetic tests presdented in this section focus on the study of non-pathological voices with a regular phonation. Although the glottal analysis of less regular voices (e.g presenting a jitter or a shimmer; or containing an additive noise component during the glottal production, as it is the case for a breathy voice) is a challenging issue, this latter problem is not addressed in the present study. 
 
\begin{table}[!ht]
\centering
\begin{tabular}{| c | c | c | c | c |}
\hline
\multicolumn{3} {| c |}{Source} & Filter & Noise\\
\hline
\hline
Pitch (Hz) & $Oq$ & $\alpha_m$ & Vowel type & SNR (dB) \\
\hline
100:5:240 & 0.3:0.05:0.9 & 0.55:0.05:0.8 & 14 vowels & 10:10:80 \\
\hline
\end{tabular}
\caption{Table of synthesis parameter variation range.}
\label{tab:Range}
\end{table}

The three source estimation techniques described in Section \ref{sec:Estimation} (CPIF, IAIF and CCD) are compared. In order to assess their decomposition quality, two objective quantitative measures are used (and the effect of noise, fundamental frequency and vocal tract variations to these measures are studied in detail in the next subsections):

\begin{itemize}

\item {\bf Error rate on NAQ and QOQ} : An error on the estimation of NAQ and QOQ after source-tract decomposition should be penalized. An example of distribution for the relative error on QOQ in clean conditions is displayed in Figure \ref{fig:Histo}. Many attributes characterizing such a histogram can be proposed to evaluate the performance of an algorithm. The one we used in our experiments is defined as the proportion of frames for which the relative error is higher than a given threshold of $\pm20\%$. The lower the error rate on the estimation of a given glottal parameter, the better the glottal flow estimation method.

\begin{figure}[!ht]
  \centering
  \includegraphics[width=0.45\textwidth]{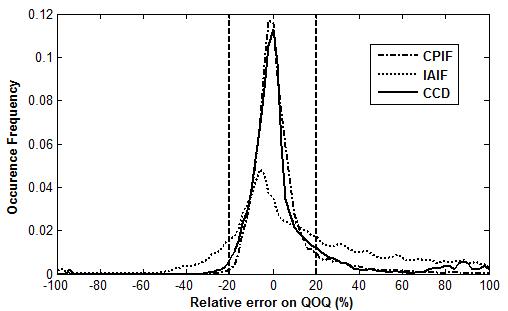}
  \caption{Distribution of the relative error on $QOQ$ for the three methods in clean conditions ($SNR = 80dB$). The \emph{error rate} is defined as the percentage of frames for which the relative error is higher than a given threshold of 20\% (indicated on the plot).}
  \label{fig:Histo}
\end{figure}

\item {\bf Spectral distortion} : Many frequency-domain measures for quantifying the distance between two speech frames $x$ and $y$ arise from
the speech coding litterature. Ideally the subjective ear sensitivity should be formalised by incorporating psychoacoustic effects such
as masking or isosonic curves. A simple and relevant measure is the spectral distortion (SD) defined as \citep{Eriksson}:

\begin{equation}\label{eq:SD}
SD(x,y) = \sqrt{\int_{-\pi}^\pi(20\log_{10}|\frac{X(\omega)}{Y(\omega)}|)^2\frac{\text{d}\omega}{2\pi}}
\end{equation}

where $X(\omega)$ and $Y(\omega)$ denote both signals spectra as a function of normalized angular frequency. In \citep{Paliwal}, authors argue that a difference of about 1dB (with a sampling rate of 8kHz) is hardly perceptible. In order to take this point into account, we used the following measure between the spectra of the estimated and reference glottal signals:
\begin{align}\label{eq:SD2}
SD(Estimated,Reference) \approx \nonumber \\
\sqrt{\frac{2}{8000}\int_{20}^{4000}{(20\log_{10}|\frac{S_{Estimated}(f)}{S_{Reference}(f)}|)^2 \text{df}}}
\end{align}
\end{itemize}

An efficient technique of glottal flow estimation is then reflected by low spectral distortion values.

\begin{figure*}[!ht]
  \centering
  \includegraphics[width=0.95\textwidth]{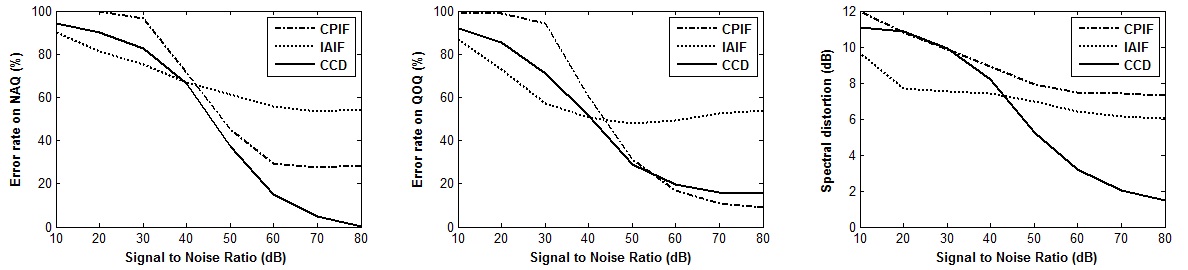}
  \caption{Evolution of the three performance measures (error rate on $NAQ$ and $QOQ$, and spectral distortion) as a function of the Signal to Noise Ratio for the three glottal source estimation methods.}
  \label{fig:SNRinfluence}
\end{figure*}

\begin{figure*}[!ht]
  \centering
  \includegraphics[width=0.95\textwidth]{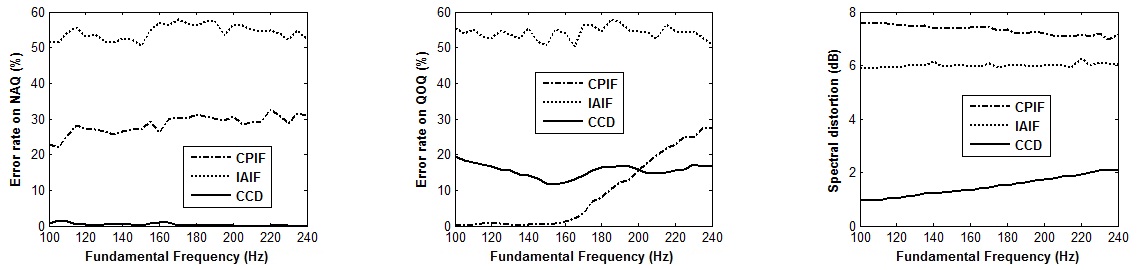}
  \caption{Evolution of the three performance measures as a function of the fundamental frequency for the three glottal source estimation methods.}
  \label{fig:PitchInfluence}
\end{figure*}

Based on this experimental framework, we now study how the glottal source estimation techniques behave in noisy conditions, or with regard to some factors affecting the decomposition quality, such as the fundamental frequency or the vocal tract transfert function.

\subsection{Robustness to Additive Noise}\label{ssec:Noise}
As mentioned above, white Gaussian noise has been added to the speech signal, with various SNR levels. This noise is used as a (weak) substitute for recording or production noise but also for every little deviation to the theoretical framework which distinguishes real and synthetic speech. Results according to our three performance measures are displayed in Figure \ref{fig:SNRinfluence}. As expected, all techniques degrade as the noise power increases. More precisely, CCD turns out to be particularly sensitive. This can be explained by the fact that a weak presence of noise may dramatically affect the phase information, and consequently the decomposition quality. The performance of CPIF is also observed to strongly degrade as the noise level increases. This is probably due to the fact that noise may dramatically modify the spectral envelope estimated during the closed phase, and the resulting estimate of the vocal tract contribution becomes erroneous. On the contrary, even though IAIF is, in average, the less efficient on clean synthetic speech, it outperforms other techniques in adverse conditions (below 40 dB of SNR). One possible explanation of its robustness is the iterative process it relies on. It can be indeed expected that, although the first iteration may be highly affected by noise (as it is the case for CPIF), the severity of the perturbation becomes weaker as the iterative procedure converges.

\subsection{Sensitivity to Fundamental Frequency}\label{ssec:PitchInfluence}
Female voices are known to be especially difficult to analyze and synthesize. The main reason for this is their high fundamental frequency which implies to process shorter glottal cycles. As a matter of fact the vocal tract response has not the time to freely return to its initial state between two glottal sollication periods (i.e. the duration of the vocal tract response can be much longer than that of the glottal closed phase). Figure \ref{fig:PitchInfluence} shows the evolution of our three performance measures with respect to the fundamental frequency in clean conditions. Interestingly, all methods maintain almost the same efficiency for high-pitched voices. Nonetheless an increase of the error rate on QOQ for CPIF, and an increase of the spectral distortion for CCD can be noticed. It can be also observed that, for clean synthetic speech, CCD gives the best results with an excellent determination of NAQ and a very low spectral distortion. Secondly, despite its high spectral errors, CPIF leads to an efficient parametrization of the glottal shape (with notably the best results for the determination of QOQ).

\subsection{Sensitivity to Vocal Tract}\label{ssec:F1Influence}

\begin{figure}[!ht]
  \centering
  \includegraphics[width=0.45\textwidth]{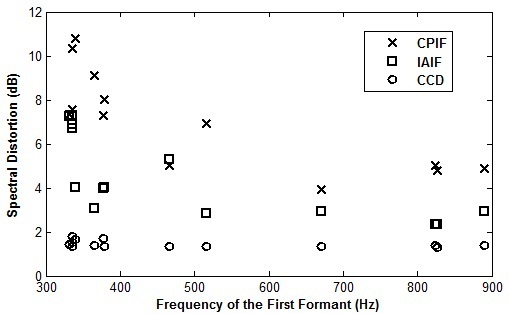}
  \caption{Evolution, for the 14 vowels, of the spectral distortion with the first formant frequency $F_1$.}
  \label{fig:F1influence}
\end{figure}

In our experiments, filter coefficients were extracted by LPC analysis on sustained vowels. Even though the whole vocal tract spectrum may affect the decomposition, the first formant, which corresponds to the dominant poles, generally imposes the longest contribution of its time response. To give an idea of its impact, Figure \ref{fig:F1influence} exhibits, for the 14 vowels, the evolution of the spectral distortion as a function of the first formant frequency $F_1$. A general trend can be noticed from this graph: it is observed for all methods that the performance of the glottal flow estimation degrades as $F_1$ decreases. This will be explained in the next Section by an increasing overlap between source and filter components, as the vocal tract impulse response gets longer. It is also noticed that this degradation is particularly important for both CPIF and IAIF methods, while the quality of CCD (which does not rely on a parametric modeling) is only slightly altered. 

\subsection{Conclusions on Synthetic Speech}\label{ssec:Factors}

Many factors may affect the quality of the source-tract separation. Intuitively, one can think about the \emph{time interference} between minimum and maximum-phase contributions, respectively related to the vocal tract and to the glottal open phase. The stronger this interference, the more important the time overlap between the minimum-phase component and the maximum-phase response of the next glottal cycle, and consequently the more difficult the decomposition. Basically, this interference is conditioned by three main parameters:

\begin{itemize}
\item the pitch $F_0$, which imposes the spacing between two successive vocal system responses,
\item the first formant $F_1$, which influences the length of the minimum-phase contribution of speech,
\item and the glottal formant $F_g$, which controls the length of the maximum-phase contribution of speech. Indeed, the glottal formant is the most important spectral feature of the glottal open phase (see the low-frequency resonance in the amplitude spectrum of the glottal flow derivative in Figure \ref{fig:MixedPhase}). It is worth noting that $F_g$ is known \citep{Doval-Spectrum} to be a function of the time-domain characteristics of the glottal open phase (i.e of the maximum-phase component of speech): the open quotient $O_q$, and the asymmetry coefficient ($\alpha_m$).
\end{itemize}

A strong interference then appears with high pitch, and with low $F_1$ and $F_g$ values. The previous experiments confirmed for all glottal source estimation techniques the performance degradation as a function of $F_0$ and $F_1$. Although we did not explicitly measure the sensitivity of these techniques to $F_g$ in this manuscript, it was confirmed in other informal experiments we performed.

It can be also observed from Figures \ref{fig:SNRinfluence} and \ref{fig:PitchInfluence} that the overall performance through an objective study on synthetic signals is the highest for the complex cepstrum-based technique. This method leads to the lowest values of spectral distortion and gives relatively high rates for the determination of both NAQ and QOQ parameters. The CPIF technique exhibits better performance in the determination of QOQ in clean conditions and especially for low-pitched speech. As for the IAIF technique, it turns out that it gives the worst results in clean synthetic speech but outperforms other approaches in adverse noisy conditions. Note that our results corroborate the conclusions drawn in \citep{Sturmel} where the mixed-phase deconvolution (achieved in that study by the ZZT method) was shown to outperform other state-of-the-art approaches of glottal flow estimation.

\section{Experiments on Real Speech}\label{sec:Real}


Reviewing the glottal flow estimation literature, one can easily notice that testing with natural speech is a real challenge. Even in very recent published works, all tests are performed only on sustained vowels. In addition, due to the unavailability of a reference for the real glottal flow (see Section \ref{sec:Intro}), the procedure of evaluation is generally limited to providing plots of glottal flow estimates, and checking visually if they are consistent with expected glottal flow models. For real speech experiments, here we will first follow this state-of-the-art experimentation (of presenting plots of estimates for a real speech example), and then extend it considerably both by extending the content of the data to a large connected speech database (including non-vowels), and extending the method to a comparative parametric analysis approach.

In this study, experiments on real speech are carried out on the De7 corpus, a diphone database designed for expressive speech synthesis \citep{Schroder}. The database contains three voice qualities (modal, soft and loud) uttered by a German female speaker, with about 50 minutes of speech available for each voice quality (leading to a total of around 2h30). Recordings sampled at 16 kHz are considered. Locations of both GCIs and GOIs are precisely determined from these signals using the algorithm described in \citep{Drugman-GCI}. As mentioned in Section \ref{sec:Estimation}, an accurate position of both events is required for an efficient CPIF technique, while the mixed-phase decomposition (as achieved by CCD) requires, among others, GCI-centered windows to exhibit correct phase properties.

\begin{figure*}[!ht]
  \centering
  \includegraphics[width=0.95\textwidth]{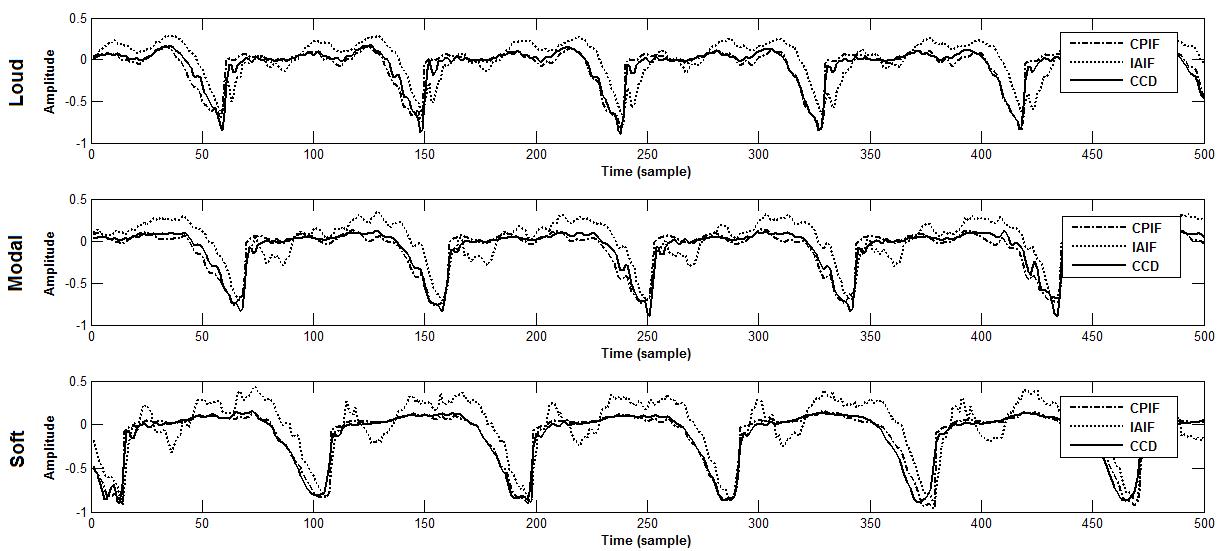}
  \caption{Example of glottal flow derivative estimation on a given segment of vowel (\emph{/aI/} as in "\emph{ice}") for the three techniques and for the three voice qualities: \emph{(top)} loud voice, \emph{(middle)} modal voice, \emph{(bottom)} soft voice.}
  \label{fig:DecompRealSpeech}
\end{figure*}

Let us first consider in Figure \ref{fig:DecompRealSpeech} a concrete example of glottal source estimation on a given voiced segment (\emph{/aI/} as in "\emph{ice}") for the three techniques and for the three voice qualities. In the IAIF estimate, some ripples are observed as if some part of the vocal tract filter contribution could not be removed. On the other hand, it can be noticed that the estimations from CPIF and CCD are highly similar and are very close to the shape expected by the glottal flow models, such as the LF model \citep{LF}. It can be also observed that the abruptness of the glottal open phase around the GCI is stronger for the loud voice, while the excitation for the softer voice is smoother.

\begin{figure*}[!ht]
  \centering
  \includegraphics[width=0.95\textwidth]{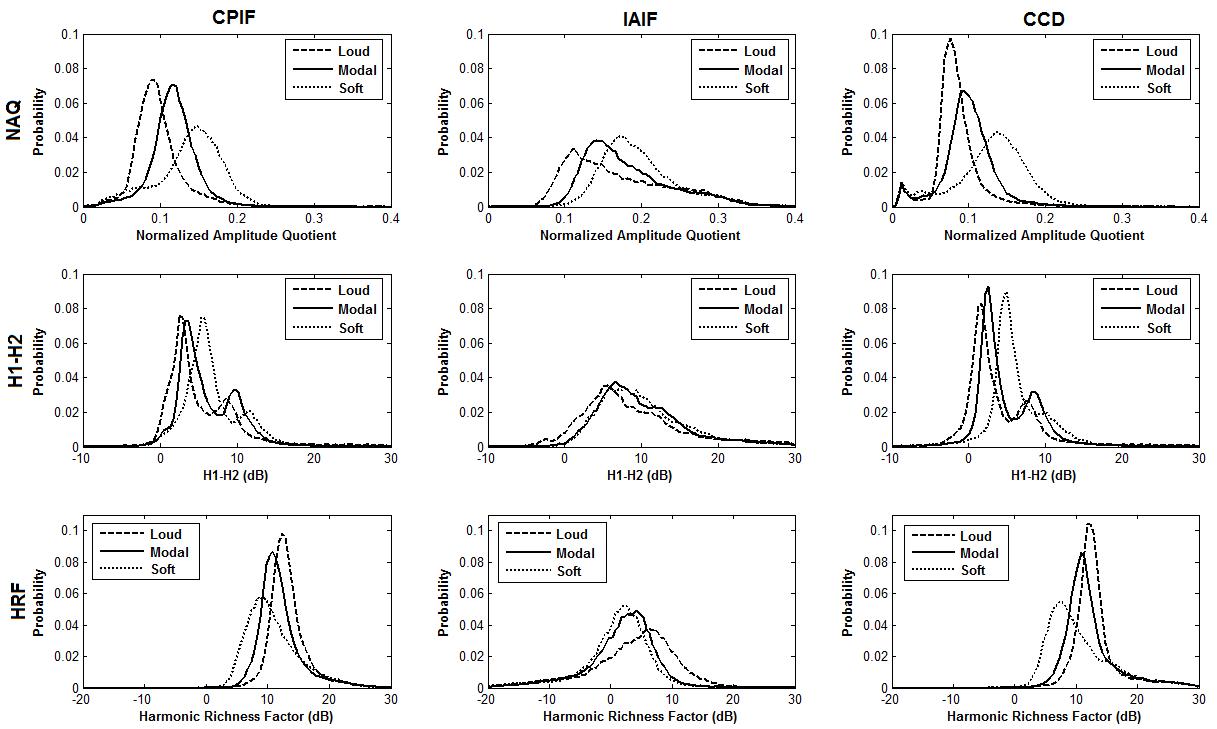}
  \caption{Distributions, for various voice qualities, of three glottal features (from top to bottom: NAQ, H1-H2 and HRF) estimated by three glottal source estimation techniques (from left to right: CPIF, IAIF and CCD). The voice qualities are shown as dashed (loud voice), solid (modal voice) and dotted (soft voice) lines.} 
  \label{fig:EmotionDistrib}
\end{figure*}

We investigated whether the glottal source estimated by these techniques conveys information about voice quality. Indeed the glottis is assumed to play an important part for the production of such expressive speech \citep{Alessandro}. As a matter of fact we found some differences between the glottal features in our experiments on the De7 database. In this experiment, the NAQ, H1-H2 and HRF parameters described in Section \ref{sec:Parametrization} are used. Figure \ref{fig:EmotionDistrib} illustrates the distributions of these features estimated by CPIF, IAIF and CCD for the three voice qualities. This Figure can be considered as a summary of the voice quality analysis using three state-of-the-art methods on a large speech database. The parameters NAQ, H1-H2 and HRF have been used frequently in the literature to label phonation types \citep{Alku-NAQ}, \citep{Hansson}, \citep{HRF}. Hence the separability of the phonation types based on these parameters can be considered as a measure of effectiveness for a particular glottal flow estimation method.

For the three methods, significant differences between the histograms of the different phonation types can be noted. This supports the claim that, by applying one of the given glottal flow estimation methods and by parametrizing the estimate with one or more of the given parameters, one can perform automatic voice quality/phonation type labeling with a much higher success rate than by random labeling. It is noticed from Figure \ref{fig:EmotionDistrib} that parameter distributions are convincingly distinct, except for the IAIF and H1-H2 combination. The sorting of the distributions with respect to vocal effort are consistent and in line with results of other works (\citep{Alku-NAQ} and \citep{Alku-CP}). Among other things, strong similarities between histograms obtained by CPIF and CCD can be observed. In all cases, it turns out that the stronger the vocal effort, the lower NAQ and H1-H2, and the higher HRF.

Although some significant differences in glottal feature distributions have been visually observed, it is interesting to quantify the discrimination between the voice qualities enabled by these features. For this, the Kullback-Leibler (KL) divergence, known to measure the separability between two discrete density functions $A$ and $B$, can be used \citep{KL}:

\begin{equation}\label{eq:DKL}
D_{KL}(A,B)=\sum_i{A(i)\log_2 \frac{A(i)}{B(i)}}
\end{equation}

\begin{figure*}[!ht]
  \centering
  \includegraphics[width=0.95\textwidth]{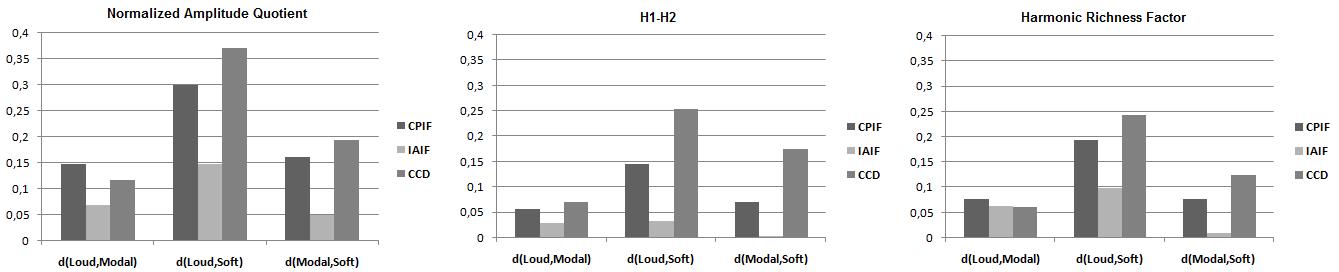}
  \caption{Jensen-Shannon distances between two types of voice quality using (from left to right) the NAQ, H1-H2 and HRF parameters. For each feature and pair of phonation types, the three techniques of glottal source estimation are compared.}
  \label{fig:Distances}
\end{figure*}

Since this measure is non-symmetric (and consequently is not a true distance), its symmetrised version, called Jensen-Shannon divergence, is often prefered. It is defined as a sum of two KL measures \citep{KL}:

\begin{equation}\label{eq:DJS}
D_{JS}(A,B)=\frac{1}{2}D_{KL}(A,M)+\frac{1}{2}D_{KL}(B,M)
\end{equation}

where $M$ is the average of the two distributions ($M=0.5*(A+B)$). Figure \ref{fig:Distances} displays the values of the Jensen-Shannon distances between two types of voice quality, for the three considered features estimated by the three techniques. 

From this figure, it can be noted that NAQ is the best discriminative feature (i.e. has the highest Jensen-Shannon distance between distributions), while H1-H2 and HRF convey a comparable amount of information for discriminating voice quality. As expected, the loud-soft distribution distances are highest compared to loud-modal and modal-soft distances. In seven cases out of nine (three different parameters and three different phonation type couples), CCD leads to the most relevant separation and in two cases (loud-modal separation with NAQ, loud-modal separation
with HRF) CPIF provides a better separation. Both Figures \ref{fig:EmotionDistrib} and \ref{fig:Distances} show that the effectiveness of CCD and CPIF is similar, with slightly better results for CCD, while IAIF exhibits clearly lower performance (except for one case: loud-modal separation with HRF).

\section{Conclusion}\label{sec:conclu}
This study aimed at comparing the effectiveness of the main state-of-the-art glottal flow estimation techniques. For this, detailed tests on both synthetic and real speech were performed. For real speech, a large corpus was used for testing, without limiting analysis to sustained vowels. Due to the unavailability of the reference glottal flow signals for real speech examples, the separability of three voice qualities was considered as a measure of the ability of the methods to discriminate different phonation types. In synthetic speech tests, objective measures were used since the original glottal flow signals were available. Our first conclusion is that the usefulness of NAQ, H1-H2 and HRF for parameterizing the glottal flow is confirmed. We also confirmed other works in the literature (such as \citep{Alku-NAQ} and \citep{Alku-CP}) showing that these parameters can be effectively used as measures for discriminating different voice qualities. Our results show that the effectiveness of CPIF and CCD appears to be similar and rather high, with a slight preference towards CCD. However, it should be emphasized here that in our real speech tests, clean signals recorded for text-to-speech (TTS) synthesis were used. We can thus confirm the effectiveness of CCD for TTS applications (such as emotional/expressive TTS). However, for applications which require the analysis of noisy signals (such as telephone applications) further testing is needed. We observed that in the synthetic speech tests, the ranking dramatically changed depending on the SNR and the robustness of CCD was observed to be rather low. IAIF has lower performance in most tests (both in synthetic and real speech tests) but shows up to be comparatively more effective in very low SNR values. 

\section*{Acknowledgment}

Thomas Drugman is supported by the Belgian Fonds National de la Recherche Scientifique (FNRS). Authors also would like to thank the reviewers for their fruitful comments.

\bibliographystyle{model2-names}
\bibliography{elsarticle-template-2-harv}








\end{document}